\documentclass[conference]{IEEEtran}
\IEEEoverridecommandlockouts
\usepackage{cite}
\usepackage[caption=false, font=footnotesize]{subfig}
\usepackage{amsmath,amssymb,amsfonts}
\usepackage{algorithmic}
\usepackage{graphicx}
\usepackage{textcomp}
\usepackage{xcolor}
\def\BibTeX{{\rm B\kern-.05em{\sc i\kern-.025em b}\kern-.08em
    T\kern-.1667em\lower.7ex\hbox{E}\kern-.125emX}}

\usepackage{fancyhdr}
\begin{document}

\title{Behind the Smile: Mental Health Implications\\ of Mother-Infant Interactions\\ Revealed Through Smile Analysis\thanks{This work is supported by the California Initiative to Advance Precision Medicine – ACES and The Saban Research Institute Developmental Neuroscience and Neurogenetics Program, CHLA.}
}

\author{\IEEEauthorblockN{A'di Dust}
\IEEEauthorblockA{\textit{Computer Science Department} \\
\textit{University of Southern California}\\
Los Angeles, USA \\
dust@usc.edu}
\and
\IEEEauthorblockN{Pat Levitt}
\IEEEauthorblockA{\textit{Children’s Hospital Los Angeles} \\
Los Angeles, USA \\
plevitt@chla.usc.edu}
\and
\IEEEauthorblockN{Maja Matarić}
\IEEEauthorblockA{\textit{Computer Science Department} \\
\textit{University of Southern California}\\
Los Angeles, USA \\
mataric@usc.edu}}

\maketitle
\thispagestyle{fancy}

\begin{abstract}
  Mothers of infants have specific demands in fostering emotional bonds with their children, characterized by dynamics that are different from adult-adult interactions, notably requiring heightened maternal emotional regulation. In this study, we analyzed maternal emotional state by modeling maternal emotion regulation reflected in smiles. The dataset comprises N=94 videos of approximately 3 ± 1-minutes, capturing free play interactions between 6 and 12-month-old infants and their mothers. Corresponding demographic details of self-reported maternal mental health provide variables for determining mothers' relations to emotions measured during free play. In this work, we employ diverse methodological approaches to explore the temporal evolution of maternal smiles. Our findings reveal a correlation between the temporal dynamics of mothers' smiles and their emotional state. Furthermore, we identify specific smile features that correlate with maternal emotional state, thereby enabling informed inferences with existing literature on general smile analysis. This study offers insights into emotional labor, defined as the management of one’s own emotions for the benefit of others, and emotion regulation entailed in mother-infant interactions.
\end{abstract}


\section{Introduction}

In recent years, there has been significant exploration into the detection of maternal mental health difficulties \cite{saqib2021machine} and the enduring implications of maternal mental health on infant caregiving \cite{kimmel2020maternal}. While the affective computing research community has conducted extensive work on the detection and prediction of mental health issues \cite{ashraf2020review},\cite{ceccarelli2022multimodal},\cite{mulay2020automatic}, there is limited literature to date that integrates such detection with maternal emotional status via multiple emotional state measures (e.g., scores of stress, depression, social support), delaying a comprehensive understanding of how potential maternal mental health difficulties may manifest. As an overarching issue, detecting and understanding maternal mental health is an important challenge, because, in addition to the impact on maternal well-being, it also can impact infant physical, cognitive, and social-emotional development \cite{oyetunji2020postpartum},\cite{christensen2020impact},\cite{aktar2019fetal}. 

Early infant parenting is universally challenging, and mothers represent the primary caregivers in most cases, taking on the majority of parental labor, on average \cite{lang2014relations}. Research is needed to address knowledge gaps in understanding the nuances of maternal mental health, which will facilitate the development of effective coping tools and strategies \cite{razurel2013relation}.

Due to intensive emotional labor involved, maintaining effort over time in longer interactions becomes more difficult \cite{hochschild2015managed}. To extend the state of affective computing research, our work employs statistical and machine learning methods on a dataset of mother-infant interactions and maternal self-reports of emotional state (as measured by various scales such as the Patient Health Questionnaire). The analyses are designed to identify and gain deeper insights into affective signals and their implications for mothers' mental health. The work focuses on the analysis of smiles, as smiling constitutes a prominent feature of caregiver-infant interactions \cite{mendes2014different}. We posit that signals from infant-directed smiles will aid in the detection and understanding of maternal mental health. Leveraging the existing literature on smile analysis, we draw conclusions about the social nature \cite{schmidt2003signal} and spontaneity of maternal smiles \cite{schmidt2009comparison}. Furthermore, we postulate that these features can be used to predict maternal mental health.

\section{Background}

\subsection{Maternal Mental Health}

The presence of depression, stress, social isolation, and other mental health factors can greatly impact a mother and their infant. Episodes of maternal depression are common, with 10\% of mothers exhibiting depressive episodes within the first year after delivery; 10\% of those mothers continue to have signs of depression after a full year after delivery \cite{cooper1995course}. Despite this, up to 50\% are undiagnosed due to concerns about privacy and stigma \cite{beck2006postpartum}. These challenges in maternal mental health have a negative impact on both maternal and infant well-being, influencing infant cognitive and social-emotional development \cite{porter2019perinatal},\cite{liu2017maternal}. 

Recognizing the prevalence and severity of maternal depression highlights the need for comprehensive research and intervention efforts. By focusing research efforts on understanding maternal depression, scalable best practices can be developed and applied to support the mental health of mothers and, consequently, safeguard the well-being and healthy development of their infants, emphasizing and supporting the interconnectedness of maternal and infant health.

Numerous studies have proposed methods for automatically detecting mental health issues using facial features \cite{laksana2017investigating},\cite{tran2021modeling},\cite{jahangir2003detection},\cite{shafiei2020identifying}. For example, in a study by Boman, Downs, Karali, and Pawlby. \cite{boman2020toward}, facial expression detection was used to model the effectiveness of stays in Mother-Baby Units (in-patient psychiatric facilities for mothers and their infants aged one year or younger) in reducing maternal psychiatric symptoms . While that work and others focus on detection and outcomes, there is a lack of exploration into the underlying mechanisms for the observed effects and the interactions between maternal mental health and mother-infant dynamics. In this work, we used an essential infant-directed facial expression, i.e., maternal smiles, to analyze the context of maternal depression and facial reactions.

\subsection{Smile Analysis}

The method of studying mothers’ emotional labor relies on analyzing smiles of mothers engaged in free play with their infants. In past research, temporal-location analysis of the mouth, eyebrows, and eyelids\cite{cohn2004automatic},\cite{cohn2004the},\cite{dibeklioglu2010eyes},\cite{geld2008tooth},\cite{park2020differences},\cite{valstar2007how},\cite{wu2014spontaneous},\cite{yao2023understanding} and detection methods such as the use of electromyography \cite{perusquiahernandez2017spontaneous},\cite{hess1988an} have enabled the prediction of the spontaneity of smiles. Moreover, past work has demonstrated that smile features can predict the social context in which the smile occurs, distinguishing between social and solitary settings \cite{fridlund1991sociality},\cite{schmidt2003signal}. Hence, smile analysis serves as a tool for decoding non-verbal communication, although not necessarily internal affective states. While traditionally applied in automatic smile and mood detection \cite{hernandez2012mood},\cite{jaques2015smiletracker},\cite{kaur2022i} and engagement analysis \cite{latulipe2011love}, our work explores a novel direction: determining whether smile analysis can provide crucial insights into the interaction and emotional labor of mothers during dyadic engagements with their infants, particularly during pivotal stages of child development.

Recent studies caution against relying solely on smiles for emotional inference \cite{barrett2019emotional}. Accordingly, we focused on correlating known contextual factors of smiles with their associations to maternal mental health at the time of audio-video data collection of mother-infant free play. Other studies report that individuals with minor depression exhibit heightened sensitivity to others' smiles \cite{gadassi2016confusing}, while also deriving benefits from their own smiles, whether posed or spontaneous \cite{lin2015helpful}. Thus, despite the potential disparity between smiling and emotional states, smiles remain significant indicators of mental health challenges, especially depression. Lwi et al. \cite{lwi2019genuine}, found that smiles perceived as genuine (i.e., Duchene smiles) can improve mental health of caregivers . Given the acknowledged benefits and significance of smiles in mental health and caregiving relationships, the smile analysis in this work provides deeper insights into mother-infant interactions.

\subsection{Mother-Infant Interaction}

The bond between mothers and infants is a cornerstone of child development, shaping infants’ social-emotional and cognitive growth. Understanding the dynamics of this relationship is key to deciphering the mechanisms that underlie experience-dependent development of infants and the specific influence of maternal emotional states. Mothers' emotional availability and responsiveness provide infants with a secure base from which to explore and develop a sense of trust, ultimately influencing their self-esteem and ability to form healthy relationships throughout their lives \cite{bohlin2000attachment},\cite{easterbrooks1979the},\cite{gerhold2002early},\cite{kochanska1999implications},\cite{2003infantmother},\cite{pastor1981the},\cite{ranson2008the}. A secure attachment between mothers and infants correlates to positive emotional \cite{bohlin2000attachment}, cognitive \cite{aviezer2002school}, physical \cite{sandra2018anxiety}, and social outcomes \cite{easterbrooks1979the},\cite{gerhold2002early},\cite{kochanska1999implications},\cite{pastor1981the} later in children's lives \cite{ranson2008the}. Mother-infant dyadic interactions promote infant exploration and gaining understanding of their social world \cite{easterbrooks1979the}. Thus, recognizing the significance of the maternal-infant bond is not only pivotal in child development but also underscores the importance of supporting the well-being of mothers in this essential role.

Central to understanding mother-infant social and emotional outcomes is the concept of {\it emotional labor}. Introduced by Hochschild, the term refers to the emotional work individuals perform to manage their expression of feelings during social interactions \cite{hochschild2015managed}. Initially, emotional labor was described in the context of traditional labor situations, particularly in woman-dominated service jobs \cite{ashforth1993emotional},\cite{grandey2000emotional},\cite{johnson2007service},\cite{scott2011a}. In recent years, emotional labor has been expanded to include unpaid labor such as work in the home \cite{dean2021the},\cite{ekman2005facial},\cite{hochschild1979emotion}. The emotional experiences of mothers caring for their infants are dynamic and include teaching their children emotional regulation techniques such as the `serve and return’ dynamic in dyads \cite{national_2012}. The term ‘serve and return’ refers to the process in which the child or caregiver initiates an interaction using vocalizations or other cues and the partner responds with a directed engagement. Emotional regulation skills from infancy to adulthood have been shown to be learned from both primary and secondary caregivers, highlighting the importance of maternal modeling of emotional labor in raising their infants \cite{frankel2012parental},\cite{gross2014handbook},\cite{halberstadt1986family},\cite{morris2007the}. By analyzing how mental health states manifest in emotional labor through the temporal analysis of smiles, this work aims to provide a greater understanding of the impact of mental health difficulties on emotional regulation of children past infancy. 

\section{Methods}

\subsection{Dataset}

The Children's Hospital of Los Angeles recorded interactions between mothers and their infants, first when the infants were 6 months old (± 1 month) and then again when the same infants were 12 months old (± 1 month), resulting in a total of 94 distinct recorded interactions under IRB CHLA-21-00174. The first set of interactions is referred to as the 6-month checkup and the second as the 12-month checkup. Each interaction session lasted approximately 3 ± 1-minutes and involved a mother and an infant involved in free play; the mother was seated across from the baby, who was either in an infant seat or on the floor. Video-audio recordings were taken from various angles with an emphasis on the head-forward camera data of the mother and infant. In this work, we utilized only the videos of the mother.

We employed OpenFace \cite{baltruvsaitis2016openface} to extract the mothers’ 2-dimensional facial features and use them to extract and analyze maternal smiles during the dyadic interaction. To ensure data integrity and eliminate frames with potential occlusions, we utilized OpenFace's frame confidence feature and included only frames with a confidence level of 80\% or higher in the analysis. This threshold was chosen empirically to balance data quality and include a sufficient number of frames. Using this threshold criterion, at the 6-month checkup, 90.6\% of frames were classified as high confidence, while at the 12-month checkup, 88.7\% were classified in the same category.

In addition to the video dataset, we had access to the mothers’ mental health and early adversity questionnaire scores for each checkup. The dataset includes mothers with notably higher negative emotional state scores than the national average, allowing for more balanced classes. The scoring was performed using the following questionnaires: 1) the Adverse Childhood Experiences (ACEs), a retrospective assessment of early adversity experienced by the adult \cite{boullier2018adverse}; 2) Social Support \cite{pascoe1982construct}, a measure of maternal social network; 3) Perceived Stress Score (PSS) \cite{scale1983perceived}; 4) Pediatric ACEs and Related Life-Events Screener (PEARLS), a current state of experiencing early adversity \cite{ye2023pediatric}; and 5) the Patient Health Questionnaire 9 (PHQ-9), which reveals depressive symptoms \cite{williams2014phq}. The scores were collected by trained research staff Children's Hospital of Los Angeles each of whom were validated by a neuropsychologist to be trained to administer and score the above-listed questionnaires collected at each visit. We created categorical variables using mental health experts’ suggested score boundaries conforming to published data for the PHQ-9 \cite{kroenke2001phq}, ACES \cite{awareace}, and PSS \cite{scale1983perceived} scale. The relevant information, including the possible score range, category breakdown, mean score per category (as a measure of spread to be compared with error metrics in modeling), and the sample mean score are all shown in Table \ref{tab:mental_health}. Each emotional state score is stable between infant ages, with ACES for example changing by an average of only 0.89 points. The unique quality and longitudinal nature of the data collection significantly enriches the value of the data. Additionally, this paper establishes methods for a much larger future analysis of more than 300 dyads of ages 6, 12, and 24 months with deep phenotyping of both mother and child. 

The population of mother-infant dyads in the study was diverse. The mothers' mean age at the infant's 6 month checkup was 32.4 (SD=5.45). The languages spoken at home included 25\% of mothers speaking Spanish only, 25\% of mothers speaking English only, 42\% of mothers speaking both English and Spanish, and 8\% of mothers speaking another language. Of the total family incomes of the participants, 42.4\% of the family incomes of the participants were below the Los Angeles average salary. The mothers had higher emotional state scores than United States average scores, allowing for better balance in classification tasks and understanding of emotional state outliers.

\begin{table*}[t]
\centering
\caption{Description of mental health scores and categories; includes categorical scoring breakdown and sample mean scores of emotional state categories for comparison of MAE metrics.}
\label{tab:mental_health}
\begin{tabular}{|c|c|l|c|c|}
\hline
\textbf{Scale} &
 \textbf{Possible Score Range} &
 \multicolumn{1}{c|}{\textbf{Categories}} &
 \textbf{Mean Points per Emotional State Category} &
 \textbf{Sample Mean} \\ \hline
PHQ-9 &
 {[}0, 27{]} &
 \begin{tabular}[c]{@{}l@{}}Minimal: {[}0-4{]}\\ Mild: {[}5-9{]}\\ Moderate: {[}10-14{]} \\ Moderately Severe: {[}15-`19{]} \\ Severe: {[}20-27{]}\end{tabular} &
 $5.6 (SD = 1.2)$ &
 $3.19 (SD = 4.16)$ \\ \hline
ACES &
 {[}0, 10{]} &
 \begin{tabular}[c]{@{}l@{}}Low Risk: {[}0{]}\\ Intermediate Risk: {[}1-3{]}\\ High Risk: {[}4-10{]}\end{tabular} &
 $3.3 (SD = 2.62)$ &
 $2.07 (SD = 2.55)$ \\ \hline
Social Support &
 {[}0, 100{]} &
 \textbf{NA} &
 NA &
 $82.80 (SD = 15.97)$ \\ \hline
PSS &
 {[}0, 40{]} &
 \begin{tabular}[c]{@{}l@{}}Low: {[}0-13{]}\\ Moderate: {[}14-26{]}\\ High: {[}27-40{]}\end{tabular} &
 $13.7 (SD = 0.47)$ &
 $12.52 (SD=6.66)$ \\ \hline
PEARLS &
 {[}0, 10{]} &
 \textbf{NA} &
 NA &
 $0.47 (SD=0.90)$ \\ \hline
\end{tabular}
\end{table*}

\subsection{Maternal Smile Feature Extraction}

To identify maternal smiles in the videos, we focused on frames in which the Facial Action Coding Guide's Action Unit 12 (AU12) \cite{ekman1978facial}, related to the lip corner puller action unit (activation of the Zygomatic Major muscle), equalled or exceeded 1.5 out of 5 at least twice in one second of consecutive high-confidence video frames. This threshold for AU12 activation was based on previous reports on smile extraction \cite{mohammed2022automated}.

To extract complete smiles, we followed established protocols for smile analysis \cite{cohn2004the},\cite{schmidt2008comparison},\cite{schmidt2003signal}, which involved identifying smile onset, apex, and offset. We standardized the coordinates of the right and left lip corners to mitigate potential motion artifacts caused by head movements. The inner left nostril coordinates were used as a reference point, and the left and right lip corner coordinates were subtracted from these central coordinates, as in \cite{schmidt2003signal}.

To determine smile phases, we calculated an initial smile radius to account for facial differences among mothers. The initial radius ($IR$) was computed as
$\frac{\sqrt{(RT_x - L_x)^2 + (RT_y - L_y)^2}}{2}$, where $RT$ and $L$ represent the coordinates of the right and left lip corners, respectively. In subsequent frames, we calculated the rightward movement $r$, where $r = \frac{\sqrt{RT_x^2 + RT_y^2}}{IR}$.

\begin{figure}
  \centering
  \includegraphics[width=0.4\textwidth]{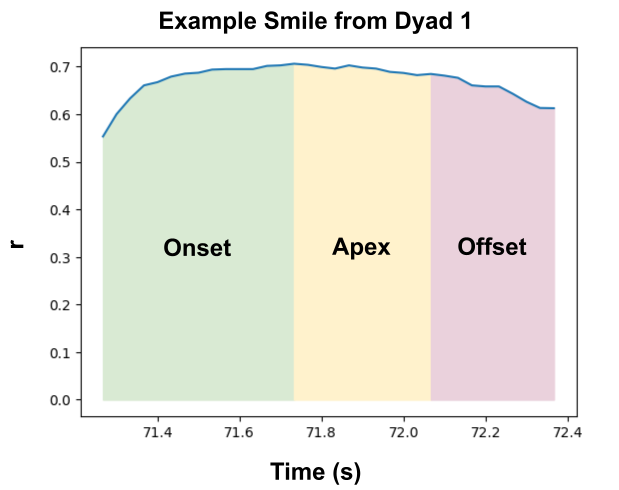}
  \caption{A graph of an example smile from the mother in dyad 1. The graph displays the smile onset, apex, and offset.}
  \label{fig:smile}
\end{figure}

The longest increasing value of r was recorded as the smile onset, while the longest decreasing value of r after the onset indicated the smile offset. The time between onset and offset represented the apex of the smile. Following the offset, the initial radius was recalculated to distinguish subsequent AU12 activations as different smiles. Figure \ref{fig:smile} shows an illustrative example of a smile with labeled onset, apex, and offset. Figure \ref{fig:smile} example of a smile devoid of large amounts of irregularities; the image on the right shows a more typical, smile with noise within the apex. Consistent with other studies \cite{schmidt2008comparison}, our modeling approach uses smile onsets and offsets instead of apexes, which tend to be more noisy.

During the 6-month checkup, mothers smiled an average of $5.02 (SD=5.02)$ times in the interaction, whereas during the 12-month checkup, mothers smiled an average of $5.30 (SD=4.45)$ times. These are adjusted values after excluding smiles confounded by speech. We employed voice activity detection using the Python library SpeechRecognition \cite{zhang2017speech} to identify segments of speech. Speech was defined as vocalizations lasting longer than 10 milliseconds (ms); vocalizations lasting less than 10 ms often lacked accompanying lip movements. Subsequently, smiles that had overlapping start and end timestamps within the range of the extracted speech were removed. This process led to the elimination of 181 potentially confounded smiles, retaining 67.6\% of the original smiles.

\subsection{Temporal Feature Windowing}

We hypothesized that variations in emotional labor are linked to the emotional status of mothers over the course of an interaction, since emotion regulation is more difficult for those with mental health difficulties \cite{Saxena_Dubey_Pandey_2011}. We used neural networks to explore the correlation between temporal aspects and maternal smile characteristics for detecting maternal mental health challenges.  The details of the networks architecture are provided in the next section.

To explore temporal patterns in maternal smiles, we tested neural networks with windows of size 1, 2, 3, 4, and 5. The choice to use windowing as opposed to recurrent neural networks was driven by the relative accuracy of windowing with small datasets and short temporal sequences 
\cite{zaremba2014recurrent}. The average number of smiles that were not confounded by speech in each session was $5.30 (SD = 4.73)$.

We calculated the mean of maternal smile features over each of these windows. Additionally, we ran each neural network with 50, 100, 150, and 200 epochs. Final neural networks were selected using the smallest mean average error. To test the stability of our results, each network was tested over 5 different randomized seeds with 5 fold cross validation and confirmed to be consistent across seeds.

\subsection{Model Architecture}

As tree-based networks perform worse at extrapolation, we chose to use neural networks. Since we are particularly interested in cases where emotional state is highly impacted and there are few of these instances in our dataset, it is important to consider extrapolation. For example, the maximum PHQ value is 22 out of 27 in our dataset and the lowest Social Support Score is 19 out of 100. Without the ability for strong extrapolation, trends after these extremes are undefinable under tree-based methods \cite{loh2007extrapolation}.

Each neural network was created to assess the temporal relation between maternal smiles and their mental health. Each measure of mental health was presented in a separate model, as we theorized the relationship between smile windowing and mental health would be dependent on questionnaire score type. We predicted differences in the temporal relation between, for example, social support and ACES, since some mental health features are more correlated than others. 

Out of the total of 94 mothers, we used 75 for model training and the remaining 19 for model testing. Model inputs included averages of smile features over windows of sequential smiles for each mother. The windowed features were passed through a fully connected (FC) layer with 32 neurons, another FC layer of 32 neurons, a final layer of 8 neurons, and linear outputs were collected through a ReLU activation function. The bottleneck layer of 8 neurons was selected so that feature redundancy could be reduced before the final activation function. The neural network used an Adam optimizer and a Mean Standard Error (MSE) loss function.

\section{Results}

\subsection{Temporal Analysis}

\begin{figure*}
  \centering
  \subfloat[]{\includegraphics[width=0.3\textwidth]{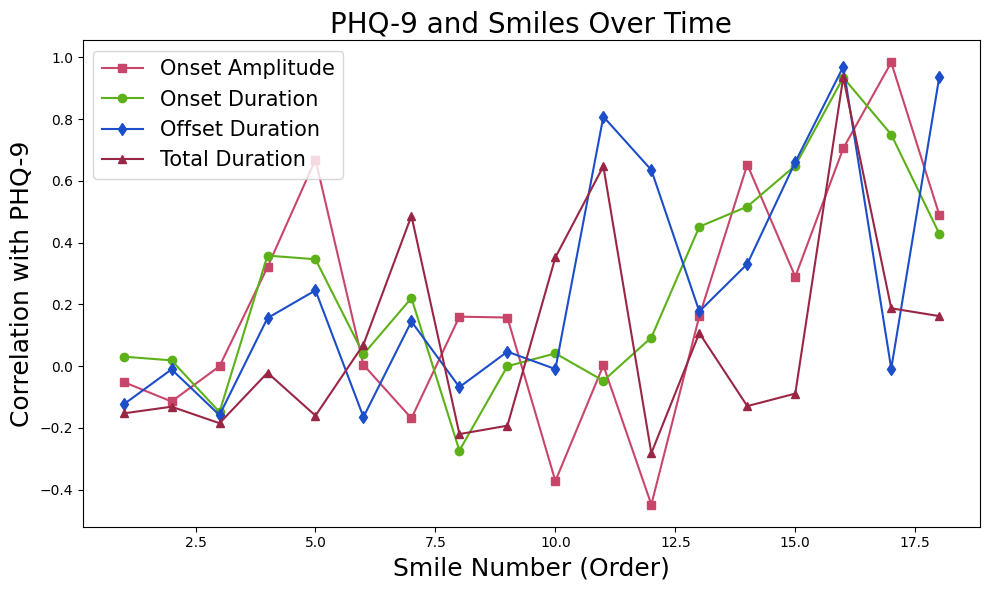}}
  \label{fig:phq_corr}
  \hfil
  \subfloat[]{\includegraphics[width=0.3\textwidth]{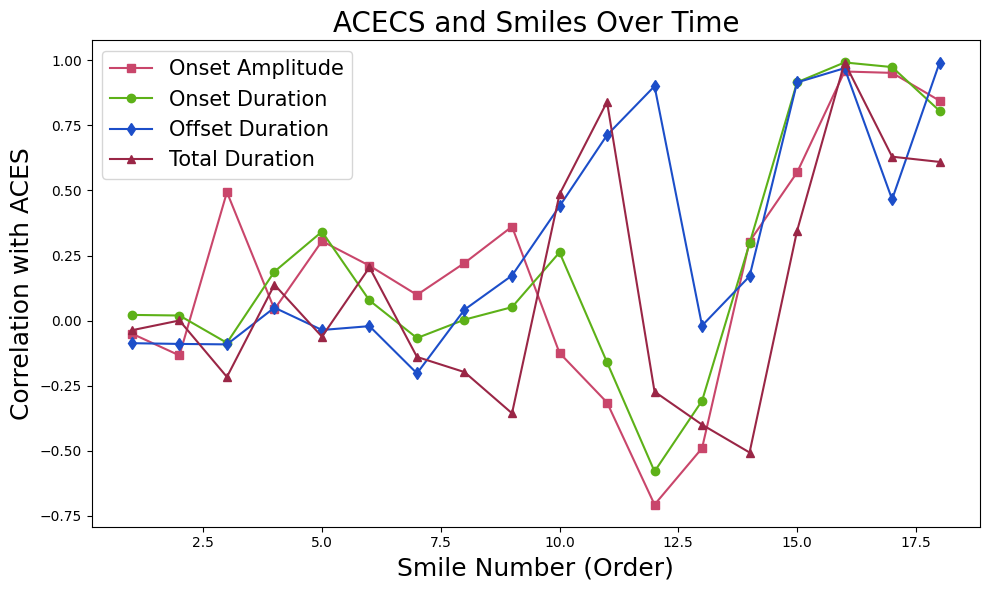}}
  \hfil
  \subfloat[]{\includegraphics[width=0.3\textwidth]{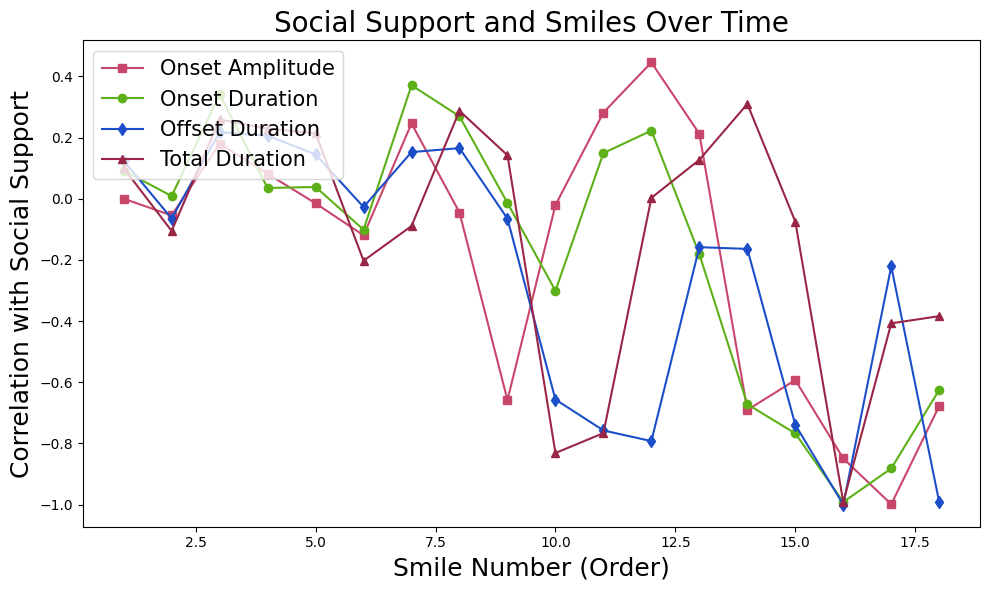}}
  \hfil
  \subfloat[]{\includegraphics[width=0.3\textwidth]{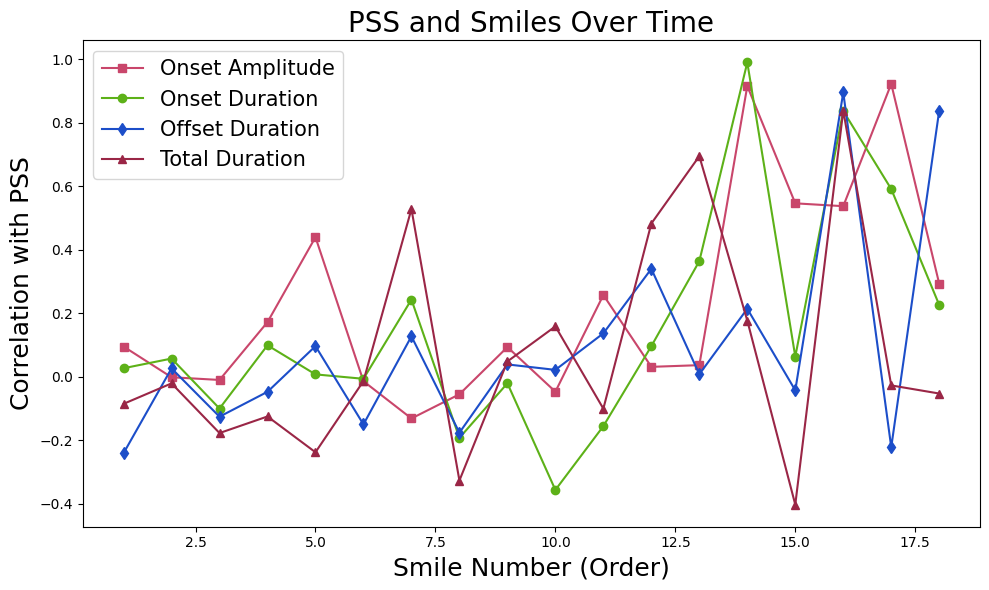}}
  \hfil
  \subfloat[]{\includegraphics[width=0.3\textwidth]{pss_corr.png}}
  \hfil
  \subfloat[]{\includegraphics[width=0.3\textwidth]{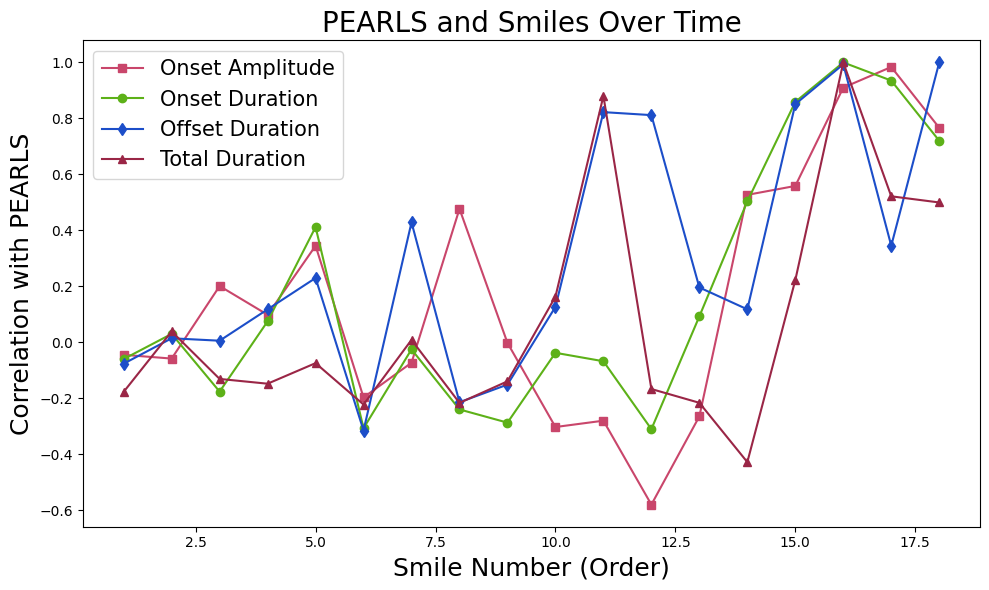}}
  \caption{The change in smile feature correlations by emotional state scale scores over each smile in the interaction. The most linearly correlated smile features were selected.}
  \label{fig: corr_graphs}
\end{figure*}

To determine if there is a relation between where in the course of the 3-minute interaction a mother’s smile is displayed and the same mother’s questionnaire scores, we first explored linear relationships between the smile features and the scores. Sequentially analyzing each smile, we calculated the sample correlation between each smile feature and each score (e.g., determining the correlation between the onset amplitude of the first smile in an interaction and the PHQ-9 score). Subsequently, we employed simple linear regression on these pairs of smile features and scores and identified the four smile features most linearly correlated with the sequential smile number and the score: onset amplitude, onset duration, offset duration, and total duration. These four features were chosen based on significance of $p < 0.05$ of their correlative relation on average. The relationships derived from these analyses are shown in Figure \ref{fig: corr_graphs}. 

The data show that, as indices of potential mental health challenges increase, so do onset amplitude, onset duration, offset duration, and total duration of smiles. Moreover, this correlation tends to intensify over time, suggesting that heightened levels of these smile features correspond to more pronounced scores as the mother-infant interaction progresses. Past research suggests that smiles characterized by higher onset amplitudes, onset durations, and offset durations are often associated with more deliberate and socially-directed expressions \cite{schmidt2006movement}. This observation implies that mothers may engage in heightened ``performance" for their infants as the duration of one-on-one interactions extends, particularly if they have greater mental health challenges. This interplay between smile features, emotional status, and the passage of time within an interaction offers insights into how emotional labor is distributed within interactions.

Recognizing the complexity and potential non-linearity of the relationship among the questionnaire scores, smile features, and time, we performed further modeling and analysis. Specifically, we trained the neural network outlined in \ref{tab:neural_net} using different window sizes and epochs. We tested windows of 2, 3, 4, and 5 smiles and 50, 100, 150, and 200 epochs. Each feature combination underwent 5 iterations with different seeds, and the final model was selected based on the minimum Mean Squared Error (MSE), averaged across the 5 seeded iterations. Given the substantial correlation between  differing scales, we developed individual networks for each questionnaire score. The resulting Mean Absolute Error (MAE) and Root Mean Squared Error (RMSE) are detailed in Table \ref{tab:neural_net}, along with the corresponding parameters: the number of epochs and the size of the sliding window.

\begin{table}[]
\centering
\caption{Best window size and number of epochs for MLP model, along with their corresponding mean absolute error and root mean squared error for each mental health score.}
\label{tab:neural_net}
\begin{tabular}{|c|c|c|c|c|}
\hline
\textbf{\begin{tabular}[c]{@{}c@{}}Metric and \\ Range of Scores\end{tabular}} &
 \textbf{\begin{tabular}[c]{@{}c@{}}Best\\ Window Size\end{tabular}} &
 \textbf{\begin{tabular}[c]{@{}c@{}}Best Number of\\ Epochs\end{tabular}} &
 \textbf{MAE} &
 \textbf{RMSE} \\ \hline
\begin{tabular}[c]{@{}c@{}}PHQ-9\\ {[}0, 27{]}\end{tabular}  & 4 & 150 & 1.43 & 1.92 \\ \hline
\begin{tabular}[c]{@{}c@{}}ACES \\ {[}0, 10{]}\end{tabular}  & 3 & 200 & 1.30 & 1.90 \\ \hline
\begin{tabular}[c]{@{}c@{}}Social Support\\ {[}0, 100{]}\end{tabular} & 3 & 200 & 12.68 & 17.34 \\ \hline
\begin{tabular}[c]{@{}c@{}}PSS \\ {[}0, 40{]}\end{tabular}  & 5 & 150 & 3.05 & 3.85 \\ \hline
\begin{tabular}[c]{@{}c@{}}PEARLS \\ {[}0, 10{]}\end{tabular}  & 3 & 200 & 0.37 & 0.61 \\ \hline
\end{tabular}
\end{table}

Capturing information about sliding windows and using those features as input into the network, we achieved high levels of accuracy. The MAE and RMSE values for PHQ-9, ACES, and PSS were smaller than the average spread of points within each categorical diagnosis, underscoring the robustness of the model in capturing high-level information. Furthermore, each model exhibited error measures that were smaller than the spread (standard deviation) of the data. Notably, for PHQ-9, PSS, and PEARLS, the model's error was less than half the size of the standard deviation. Optimal window sizes typically ranged between 3 and 5 smiles, mirroring the visual trends depicted in Figure \ref{fig: corr_graphs}, where relative trends persisted for sequences of 3 to 5 consecutive smiles.

\subsection{Feature Correlation and Developmental Stage}

In addition to capturing information about mothers’ emotional state and experiences regarding early adversity in her own childhood (as captured in the ACES scale), maternal smiles correlated to infant developmental stages in a way that informs how infant age may influence maternal behavior. The single modality emphasis on smiles is novel and grounded in the ability of smiles to decode the social or solitary and spontaneous of posed nature of interactions. In order to examine these relationships, we performed unequal variation t-tests on each smile feature between the 6-month appointment and the 12-month appointment, as shown in Table~\ref{tab:t-tests}. All variation ratios between the 6 and 12 month features were $<$1, so we used the unequal variation t-test. The features were also non-normal, but the degrees of freedom were $>$100, so we were able to perform t-tests due to the central limit theorem \cite{kwak2017central}. 

\begin{table}[]
\caption{Unequal variation t-tests of mother smile features during 6 and 12-month infant age appointments.}
\label{tab:t-tests}
\resizebox{\columnwidth}{!}{%
\begin{tabular}{|c|c|c|}
\hline
\textbf{Smile Feature} & \textbf{Unequal Variation T-Test}  & \textbf{Effect Size (Cohen’s D)} \\ \hline
Maximum Onset Speed  & $t(207.33) = 4.37$, $p  < 0.001$ & 0.51  \\ \hline
Maximum Offset Speed  & $t(190.17) = 3.02$,$p = 0.003$  & 0.37  \\ \hline
Onset Amplitude  & $t(190.86) = 3.97$, $p  < 0.001$ & 0.49  \\ \hline
Offset Amplitude  & $t(182.07) = 2.26$, $p = 0.025$  & 0.29  \\ \hline
Onset Duration  & $t(354.53) = 0.50$, $p = 0.62$  & 0.05  \\ \hline
Offset Duration  & $t(314.51) = 0.53$, $p = 0.60$  & 0.05  \\ \hline
Apex Duration  & $t(246.39) = 1.75$, $p = 0.081$  & 0.19  \\ \hline
Total Duration  & $t(271.39) = 2.17$, $p = 0.031$  & 0.22  \\ \hline
\end{tabular}%
}
\end{table}

The results of these t-tests allowed us to reject the null hypothesis that mothers' smile features were equal at the 6-month and 12-month appointments for maximum onset speed, maximum offset speed, onset amplitude, offset amplitude, and total duration. Each of these features was smaller in the 6-month visits than the 12-month visits. Specifically, the maximum onset speed and onset amplitude exhibited a medium effect size, while the maximum onset speed, offset amplitude, and total duration demonstrated a small but non-negligible effect size. Given that higher maximum onset speed and onset amplitude are associated with posed and socially-directed smiles \cite{schmidt2003signal},\cite{schmidt2006movement}, this suggests a positive correlation between social and posed smiles for mothers of 12-month infants compared to those of 6-month-old infants. Infants between 6 and 12 months of age typically undergo developmental maturation related to attachment, social interaction, and communication, potentially eliciting a more social response from mothers, closer to interactions in social contexts with adults \cite{schmidt2003signal}.

To further link these findings with maternal emotional state and potential mental health disturbances, we computed ANOVA scores of mothers' smiles, comparing categorical questionnaire scores with infant age and smile features, identifying cases where the null hypothesis for differences in samples could be rejected. For instance, we utilized ANOVA to determine if there were variations between low, moderate, and high PSS scores in instances in which a 6-month-old infant uses the mother's smile maximum onset speed. Our findings are detailed in Table \ref{tab:anova}. 

\begin{table}[]
\caption{ANOVA analysis on categorical mental health for each infant age and mother smile feature. Significant results at $p < 0.05$ are reported.}
\label{tab:anova}
\resizebox{\columnwidth}{!}{%
\begin{tabular}{|c|c|c|l|c|}
\hline
\textbf{Smile Feature} & \textbf{Month} & \textbf{Mental Health Variable} & \multicolumn{1}{c|}{\textbf{ANOVA}} & \textbf{Partial $\eta^2$} \\ \hline
Max Onset Speed  & 6  & PHQ  & $F(3, 178) = {[}3.39{]}$, $p = 0.019$  & 0.06  \\ \hline
Offset Amplitude  & 12  & PHQ-9  & $F(4, 343) = {[}2.49{]}$, $p = 0.043$  & 0.03  \\ \hline
Max Onset Speed  & 6  & PSS  & $F(2, 179) = {[}5.31{]}$, $p = 0.006$  & 0.06  \\ \hline
Max Offset Speed  & 12  & PSS  & $F(2, 345) = {[}4.08{]}$, $p = 0.018$  & 0.02  \\ \hline
Offset Amplitude  & 12  & PSS  & $F(2, 345) = {[}3.95{]}$, $p = 0.021$  & 0.02  \\ \hline
Onset Duration  & 12  & ACES  & $F(2, 345) = {[}3.21{]}$, $p = 0.042$  & 0.02  \\ \hline
\end{tabular}%
}
\end{table}

Our results reveal potential interactions between maternal mental health, infant developmental stage, and smile features. Notably, in the assessment of PHQ-9 categories, there were small yet noteworthy differences in mothers' smile offset amplitude when their infants were 12 months old. The PSS screener categorical score exhibited the highest number of features with statistically significant between-group differences. Specifically, mothers of 12-month-old infants displayed differences in PSS categories based on their maximum offset speed and offset amplitude, albeit with a small effect size. Conversely, mothers of 6-month-old infants exhibited PSS categorical differences in maximum onset speed, characterized by a medium effect size. Lastly, differences in onset duration for mothers of 12-month-old infants were evident across categorical ACEs scores. These results suggest that infant developmental stage, mother smile features, and mental health interact with one another with potential functional implications. 

\section{Discussion}

The utilization of infant-directed smiles in predicting maternal emotional state and potential mental health challenges offers a dual benefit: maternal smile features provide insights not only into the mother's perception of an interaction but also into the emotional labor entailed in that interaction, particularly in connection with mental health challenges that may change over time. This research provides promising insights into the interaction between mother-infant play and maternal emotional state by analyzing maternal smiles contextualized in time, along with the infant developmental stage. Our findings support the hypothesis that maternal mental health correlates with changes in maternal smiles during interactions with infants over time, as evidenced by both linear modeling and neural network analyses. Additionally, we explored the influence of infant developmental stage by examining differences in smile features across infant ages and subsequently integrated questionnaires designed to capture maternal emotional state and childhood adversity experiences (ACES).

The temporal analyses revealed support for a linear correlation between smile feature and emotional labor changing direction over time. Notably, the directionality of this correlation indicated that the smiles of mothers facing greater mental health challenges became more positively correlated with features associated with social and posed expressions over the course of the mother-infant interaction. The sliding window neural network results further reinforce our finding that time and smile features can serve as predictors of maternal emotional state and potential mental health disturbances, as evidenced by the network's accurate predictions when analyzing windows spanning 3 to 5 smiles.

Our developmental findings support the premise that as infants become more socially and emotionally-aware, and their skills in verbal and non-verbal communication begin to mature, a mother’s smile becomes more posed and socially-directed. Additionally, we show that there exist correlative interactions between infant developmental stage, maternal smile features, and emotional state based on the questionnaire datasets. This suggests that maternal mental health and maternal smile features likely vary across infant developmental stages.

The reported study results have limitations. The short duration of the interaction videos, approximately 3-minutes each, limited the collection of high-confidence smiles outside of instances coinciding with maternal speech, constraining our temporal analysis to brief interaction periods. We note, however, that successful maternal coaching interventions target the ‘serve and return’ caregiver-child interactions that are episodic and brief in nature. Additionally, the collection of data in a room equipped with two cameras introduces the possibility of smile-context confounds due to the Hawthorne effect \cite{sedgwick2015understanding}. Furthermore, the cultural context of the study, confined to a single metropolitan area in the United States, may influence certain effects due to geographic and cultural nuances.

\section{Conclusion}

This paper presents trends in maternal emotional state over the course of a mother-infant interactions by analyzing the mothers' smiles. We explored how temporal trends in smiles are predictive of maternal emotional state, as well as how the developmental stage of the infant (6 or 12 months) and emotional state of the mother affect individual smiles. We find ample evidence that maternal emotional state is correlated to maternal smiles in interactions and that this correlation differs based on the age of the infant.

Future research may explore longer mother-infant interactions, the relation between maternal emotional labor and data from direct infant developmental assessments, and may capture interactions in more naturalistic interactions (e.g., at home), and at later stages of development (e.g., with toddlers). Additionally, incorporating the collection of prosody, pragmatic speech features, postural information, and other multimodal data would enrich the findings. We used smile features validated in past literature; however, multimodal models might capture more complex relationships between infant-directed behaviors and maternal emotional state and potential mental health challenges. Furthermore, future studies would benefit from incorporating measures of maternal sociality and spontaneity beyond the existing previously validated smile features we used. 

Our research provides a foundation for determining maternal mental health in the context of infant developmental age and milestone status with facial features and creates new opportunities for further exploration of the relations between mother-infant interactions and maternal mental health.

\section*{Ethical Impact Statement}

The methods utilized in this research received approval from the Children's Hospital of Los Angeles Institutional Review Board (IRB) under protocol CHLA-21-00174. Adult participants involved in the original study provided informed consent and were compensated for their time. All data were securely maintained with HIPAA compliance to protect participant privacy. Due to the potentially identifiable and medically sensitive nature of the information within this dataset, we are unable to release it to the public. However, all participants provided consent for the use of this dataset by the appropriate institutions.

The findings of this study offer potential insights into maternal mental health. While this information could contribute to the safeguarding and comprehension of maternal mental well-being, it should not be utilized for surveillance or diagnostic purposes. Consequently, we refrain from sharing our predictive models to prevent their use as diagnostic tools. However, to facilitate reproducibility, we provide detailed descriptions of the tuning and hyperparameters applied in all our models.

Moreover, while there is evidence suggesting that some smile features in posed/spontaneous or social/solitary contexts may be cross-cultural, it is also evident that certain features may be culturally dependent \cite{fang2019unmasking}. Therefore, we advise exercising caution when applying the findings of this research to contexts outside the United States.

\bibliographystyle{IEEEtran}
\bibliography{bib}

\vspace{12pt}

\end{document}